\documentstyle[12pt,epsfig]{article}

\begin{document}

\title{A neutron halo in $^8He$}
\author{A. V. Nesterov, V. S. Vasilevsky, O. F. Chernov \\
Bogolyubov Institute for Theoretical Physics, \\
Ukrainian Academy of Sciences, 252143, Kiev 143, Ukraine}
\date{\today}
\maketitle

\abstract{
 The structure of  $^8$He is investigated within a three-cluster
 microscopic model. The three-cluster configuration  $\alpha+^2n+^2n$
 was used to describe the properties of the ground state  of the nucleus.
The obtained results evidently  indicate the existence of a neutron halo in $^8$He.
}

\section{Introduction}

The development of the experimental technique made it possible to
investigate light nuclei with large neutron excess, i.e., the nuclei for
which the ratio $\eta =(N-Z)/A$ is significantly larger than for common
ones. Such nuclei lie near the drip line and are $\beta $-unstable. They
live a short time and transform by emitting electrons into nuclei with
approximately equal number of protons and neutrons. A number of unexpected
properties were discovered in those nuclei, for instance, a neutron halo. It
is natural that many attempts were undertook to explain those properties
within microscopic and semi-microscopic methods \cite{hofman97, varga+94,
Karatagl+98, Navratil+barrett98, zhuk93, kn:Dani91, csoto93, vargaR94}.

Our aim is to investigate the structure of the $^{8}$He ground state. It is
interesting that $^{8}$He has the largest value of $\eta =0.5$ among other
nucleon-stable nuclei. Note that the average values of $\eta =0.4$ for
nuclei near the neutron drip line. As early as in 1960, Ya.~B.~Zeldovich 
\cite{zeldovich60} and V.~I.~Goldansky \cite{goldansky60} indicated a
possibility of the existence of $^{8}$He isotope. It was experimentally
confirmed \cite{cerny66} in the middle of sixties. The subsequent analysis
shows that the lowest threshold of $^{8}$He decay is $^{6}$He$+2n$ and lies
2.1 MeV above the ground state and energy of the threshold $\alpha +4n$
equals 3.10 MeV (see, for instance, \cite{kn:Ajze88}).

The most complete information on light nuclei with neutron excess can
obtained with a microscopic model. In this case, the problem is connected
with solving the many-particle Schr\"{o}dinger equation with a fixed
(chosen) nucleon-nucleon interaction. The equation has to be solved with
some simplification based on one or other physical considerations. The
Resonating Group Method \cite{kn:wilderm_eng} or its Algebraic Version \cite
{kn:Fil_Okhr, kn:Fil81} is one of such methods.

In this paper, we make use of the Algebraic Version of the Resonating Group
Method in which $^{8}$He is considered as a three-cluster configuration $%
\alpha +^{2}n+^{2}n$. It is obvious that we make {\it a priori} some
assumptions on the structure of the nucleus. First, wave functions of each
cluster are modelled by the shell-model functions. Second, valent neutrons
unite in dineutron clusters. As a justification for such an assumption can
be served the fact, indicated by A. B.~Migdal \cite{migdal72}, that the
interaction between two neutrons may be increased significantly in the
presence of the third particle. It can give rise to the creation of the
dineutron clusters on the surface of a nucleus. The chosen clusterization
allow us to consider an $\alpha $-particle as a core, despite that the
lowest threshold of $^{8}$He decay is $^{6}$He$+n+n$. Earlier, A. A.
Ogloblin \cite{Ogloblin91} indicated the importance of a cluster
configuration $\alpha +4n$. He pointed out that the bound energy of two
neutrons in $^{8}$He is two times larger than that in $^{6}$He. This fact
led him to the conclusion that $^{6}$He cannot serve as a core and the
neutron halo in $^{8}$He has to be consisted of four neutrons.

Note also that the usage of dineutron clusters is a quite grounded
approximation. For example, in \cite{kn:Vasi89}, dineutron and also diproton
clusters were successfully used to describe exit channels of the reactions $%
^{3}$H$+^{3}$H$\rightarrow ^{4}$He$+n+n$ and $^{3}$He$+^{3}$He$\rightarrow
^{4}$He$+p+p$ respectively. Besides, in \cite{hansen87}, main features of $%
^{11}$Li was reproduced within the cluster configuration $^{9}$Li$+^{2}n$
with a pointless dineutron.

\section{Method}

The present method for investigation of the $^{8}$He ground state is based
on the Algebraic Version of the Resonating Group Method (AV\ RGM). For a
long time, this version was used for studying the bound states of
two-cluster systems, reactions with a few open channels, interaction of
these channels with collective monopole and quadrupole modes, and also
processes of full disintegration of light nuclei \cite{kn:fil_vasil85,
nest_8be, nest_psh}. Recently, the AV RGM was actively applied to describe
three-cluster systems \cite{kn:Vasil96, kn:Vasil97, filipp199, filipp95}.

The Algebraic Version of the Resonating Group Method is based on the usage
of an oscillator basis for solving bound state problems and problems of
continuous spectrum states. This is achieved by expanding a wave function of
inter-cluster motion in the oscillator basis. As a result, a trial
three-cluster function takes the form 
\begin{equation}
\Psi (A)=\sum_{n}C_{n}\hat{A}\left[ \Phi _{1}(A_{1})\Phi _{2}(A_{2})\Phi
_{3}(A_{3})f_{n}({\bf q}_{1},{\bf q}_{2})\right] ,  \label{eq:1}
\end{equation}
where $\hat{A}$ is the antisymmetrization operator, $\Phi _{i}(A_{i})$ are
the internal functions of the cluster, which are selected in one or other
form prior to solving the problem (for instance, in the form of
many-particle oscillator shell functions as in our case); the set of
coefficients $C_{n}$ is nothing else but a wave function in the oscillator
representation. This function should be obtained from a system of linear
equations: 
\begin{equation}
\sum_{n^{\prime }}\left[ <n,\mid \hat{H}\mid n^{\prime }>-E<n\mid n^{\prime
}>\right] C_{n^{\prime }}=0,  \label{eq:2}
\end{equation}
which is derived directly from the many-particle Schr\"{o}dinger equation.
The oscillator functions $f_{n}({\bf q}_{1},{\bf q}_{2})$, where ${\bf q}%
_{1} $ and ${\bf q}_{2}$ are Jacobi vectors fixing a position of clusters in
space, are determinate in the six-dimensional space and constitute the
irreducible representation $[N00]$ of the unitary group $U(6)$. Thus, the
composite index $n$ consists of indices (six in total) of the irreducible
representation of the $U(6)$ group and its subgroups.

The choice of one or other reductions of the $U(6)$ group is dictated by
considerations of physical lucidity and simplicity of numerical realizations
as well. To consider the bound state problem, it is convenient to use bases,
whose classification is connected with the following reduction of the $U(6)$
group: 
\[
\begin{array}{ccccccc}
U(6)\supset & U(3) & \otimes & U(3) &  & \ \ \Longrightarrow \ \  & \mid
N_{1}l_{1},N_{2},l_{2},\,LM> \\ 
& \bigcup &  & \bigcup &  &  &  \\ 
& SO(3) & \otimes & SO(3) & \supset SO(3) &  & 
\end{array}
\]
\[
\begin{array}{cccccc}
U(6)\supset & SU(3) & \otimes & U(2) & \ \ \ \ \ \ \ \ \ \ \ \ \ \
\Longrightarrow & \mid (\lambda \mu )\nu ,\,\omega LM> \\ 
& \bigcup &  & \bigcup &  &  \\ 
& SO(3) &  & O(3) &  & 
\end{array}
\]
%
%
%

The first basis is usually called the basis of two uncoupled oscillators or
bioscillator basis (BO). Each of the $SU(3)$ groups, associated with one of
the Jacobi vectors ${\bf q}_{1}$ and ${\bf q}_{2}$, generates the quantum 
numbers $N_{1}$, $l_{1}$ and $N_{2}$, $l_{2}$. They are the principal
quantum number (or the number of oscillator quanta) and partial angular
momentum along the respective Jacobi vector: 
\[
|N_{1},l_{1},N_{2},l_{2};LM> 
\]

The second basis is an ''$SU(3)"$ basis. Wave functions of this basis are
classified through\ the well-known Elliott indices ($\lambda $,$\mu $) of
the $SU(3)$ group, multiplicity index $\omega $ arising in the reduction $%
SU(3)\subset SO(3)$, and quantum number $\nu =\frac{1}{2}(N_{1}-N_{2})$
connected with oscillator quanta along the Jacobi vectors $\vec{q}_{1}$and $%
\vec{q}_{2}$: 
\[
|(\lambda \mu )\nu ;\omega LM> 
\]

The total number of oscillator quanta equals $N=N_{1}+N_{2}=\lambda +2\mu $
and defines the irreducible representation of the $U(6)$ group. For a given $%
N$, i.e., for a fixed oscillator shell, functions of both bases related to
each other through a unitary transformation, because these bases are
eigenfunctions of the same oscillator hamiltonian in the six-dimensional
space. Thus, they are equivalent. Note that the unitary matrix connecting
these two bases consists of the Clebsch-Gordan coefficients of the $SU(3)$
group for the decomposition of the product $\left( N_{1}0\right) \otimes
\left( N_{2}0\right) \Rightarrow (\lambda \mu )$. Thus 
\[
|(\lambda \mu )\nu ;\omega LM>=\sum_{l_{1},l_{2}}U\left(
N_{1},l_{1},N_{2},l_{2};(\lambda \mu )\nu ;\omega \right)
|N_{1},l_{1},N_{2},l_{2};LM> 
\]

However we make use of two bases. This is because the bioscillator basis has
more natural quantum numbers. Meanwhile, the $SU(3)$ basis is more
convenient for numerical implementation, in particular, for eliminating
Pauli-forbidden states. Besides, the usage of two bases gives additional
information on optimal subspaces, which allow one to obtain reliable results
with minimal effort.

The elimination of Pauli-forbidden states is performed by diagonalization of
the matrix of the antisymmetrization operator 
\begin{equation}
||<n\mid n^{\prime }>||,
\end{equation}
calculated between the basis functions (\ref{eq:1}). Pauli-forbidden states
correspond to those eigenfunctions of the matrix $||<n\mid n^{\prime }>||$ $%
=||<n\left| \widehat{A}\right| n^{\prime }>||$ which have zero eigenvalues.
Pauli-allowed states are a combination of original basis functions of a
given oscillator shell which are eigenfunctions of the antisymmetrization
operator. It should be noted that the matrix $||<n\mid n^{\prime }>||$ has a
block structure. Non-zero matrix elements correspond to overlapping basis
functions of the same oscillator shell, i.e. those oscillator functions
which obey the condition $N=N^{\prime }$. To solve the Schr\"{o}dinger
equation in matrix form, one has to eliminate Pauli-forbidden states. Let us 
$e_{\alpha }$ and $\{U_{n}^{\alpha }\}$ be respectively eigenvalue and
eigenfunction of the antisymmetrization operator. Then, the system of
equations (\ref{eq:2}) should be transformed to the representation of
Pauli-allowed states: 
\begin{equation}
\sum_{\alpha ^{\prime }}\left[ <\alpha \mid \hat{H}\mid \alpha ^{\prime
}>-E\delta _{\alpha ,\alpha ^{\prime }}\right] C_{\alpha ^{\prime }}=0,
\label{eq:8}
\end{equation}
where $\left\| <\alpha \mid \hat{H}\mid \alpha ^{\prime }>\right\| $ is a
matrix of hamiltonian between Pauli-allowed states connected with the matrix 
$||<n\left| \widehat{H}\right| n^{\prime }>||$ by the relation 
\[
\left\langle \alpha \mid \hat{H}\mid \alpha ^{\prime }\right\rangle
=\sum_{n,n^{\prime }}U_{n}^{\alpha }\left\langle n\left| \widehat{H}\right|
n^{\prime }\right\rangle U_{n^{\prime }}^{\alpha ^{\prime }} 
\]

In this connection, for the bioscillator basis the original scheme of
classification is totally changed, but the quantum numbers $\left( \lambda
\mu \right) $ are preserved for the $SU(3)$ basis, because the matrix $%
||<n\mid n^{\prime }>||$ is off-diagonal with respect to the quantum number $%
\nu $ only.

We omit all details of matrix elements calculations of the microscopic
hamiltonian and antisymmetrization operator, by referring reader to the
paper \cite{kn:Nester93} where one can find basic formulae and recurrence
relations for matrix elements of operators of the physical importance \
between bioscillator functions.

\section{Results}

Results, represented in this chapter, were obtained with the Volkov
potential \cite{kn:Volk65}. The only free parameter, oscillator radius $%
r_{0} $, was chosen to minimize the threshold energy of the $^{8}$He decay
into $^{4}$He and two dineutrons. It turns out to be 1.51 fm. Under such
conditions, the energy of the $^{4}$He$+^{2}n+^{2}n$ threshold equals \
-22.15 MeV and\ the bound state energy of $\alpha $-particle is -26.84 MeV.
Coulomb interaction was neglected because it leads to a shift of the bound
state energy and threshold energy by the same value.

In what follows, we use two different trees of Jacobi vectors. In the first
tree which we call ''$T$''-tree, the vector ${\bf q}_{1}$ defines the
distance between two dineutrons, and the vector ${\bf q}_{2}$ fixes the
distance between the center of mass of two dineutrons and $\alpha $%
-particle. The second tree is called $Y$-tree. In this tree, the first
vector ${\bf q}_{1}$ determines the distance between $\alpha $-particle and
one of the dineutrons, and the second vector ${\bf q}_{2}$ is connected with
the distance between the second dineutron and center of mass of the first
dineutron and $\alpha $-particle.

As we concern with the ground state only, then we need to use wave function
of the $S$-state. In this case oscillator basis is reduced significantly.
For instance, the bioscillator basis \ involves oscillator functions with
even values of $N_{1}$ and $N_{2}$. Besides, partial angular momentum $%
l_{1}=l_{2}$. Actually, we need only three quantum numbers to classify basis
functions with $L=0$. They are ($N_{1},N_{2},l=l_{1}=l_{2}$) for
bioscillator basis, and ($\lambda \mu ,\nu $) for SU(3) basis.

\begin{figure}[tbp]
\centerline{\psfig{figure=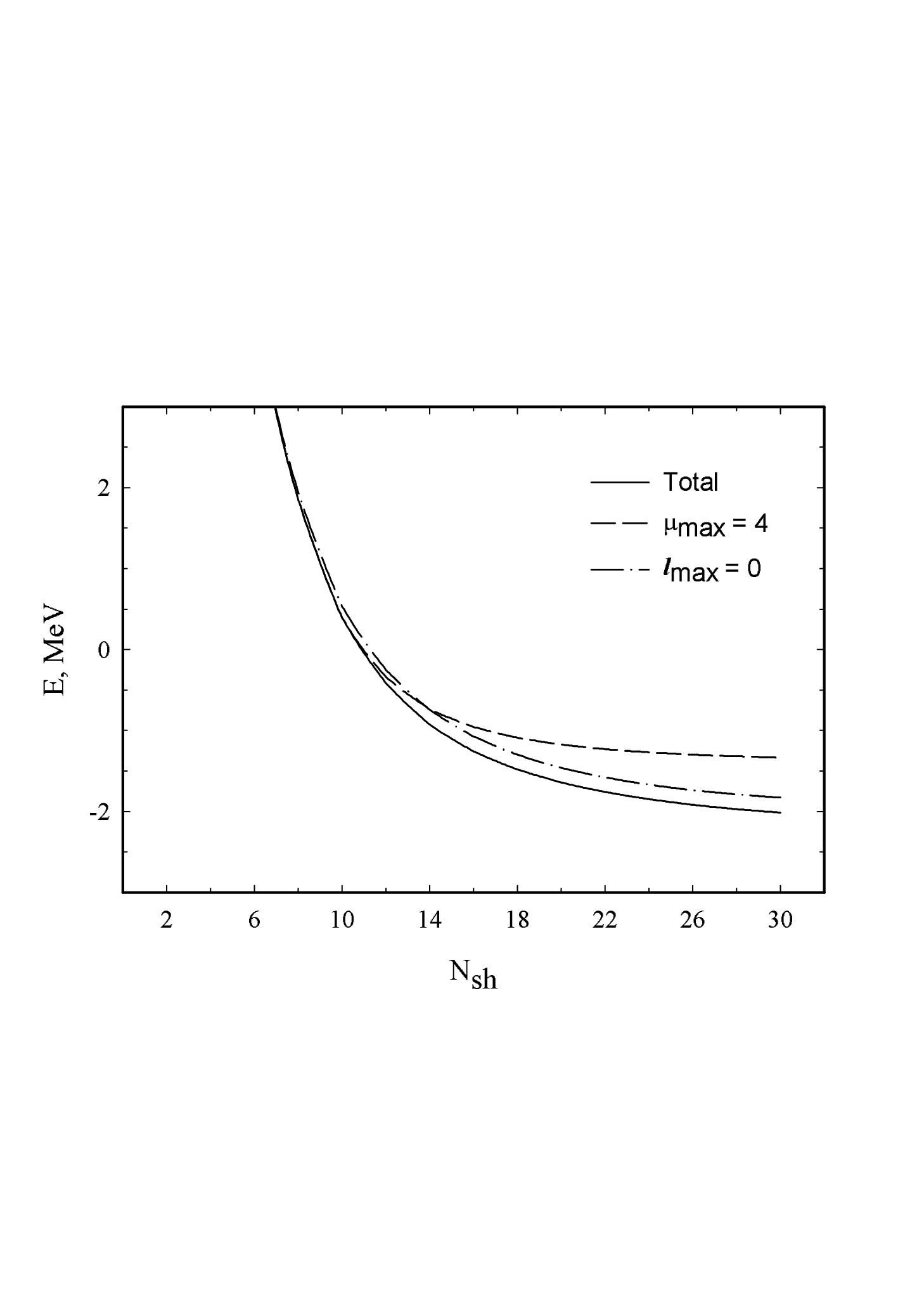}}
\caption{$^{8}$He ground state energy as
a function of the number $N$ of oscillator shells involved in the
calculation.}
\label{fig:He8_spct}
\end{figure}

{\bf Bound state energy and optimal subspaces}. The ground state of $^{8}$He
is considered with the basis which involves all oscillator functions of 15
lowest oscillator shells, i.e., basis functions with even values of the
principal quantum number $N$ up to $N=30$. The total number of the original
basis functions equals 815 and the total number of Pauli-allowed states
reduces to 399 functions. Such a number of basis functions provides a fairly
good convergence of the bound state energy, as is demonstrated in Fig. \ref
{fig:He8_spct}. In this figure, we display the ground state energy as a
function of the principal quantum number $N$. The energy is counted from the
threshold $\alpha +^{2}n+^{2}n$. In Fig. \ref{fig:He8_spct}, we also display
the ground state energy obtained with some subspaces of the total space of
the oscillator basis used. In the BO basis, such a subspace is defined by
the maximal value of the partial angular momentum $l=0$, while, for $SU(3)$%
-basis, such a subspace involves basis functions with $\mu \leq 4$. 
The
later subspace consisting of 274 functions gives the energy which noticeable
differs from ''exact'' one, obtained with the total basis. But, with the
former subspace including only 118 functions, we obtain the energy which is
very close to the ''exact'' \ value. This is probably connected with that
the interaction between clusters is most strong in the $S$-state.

\begin{figure}[tbp]
\centerline{\psfig{figure=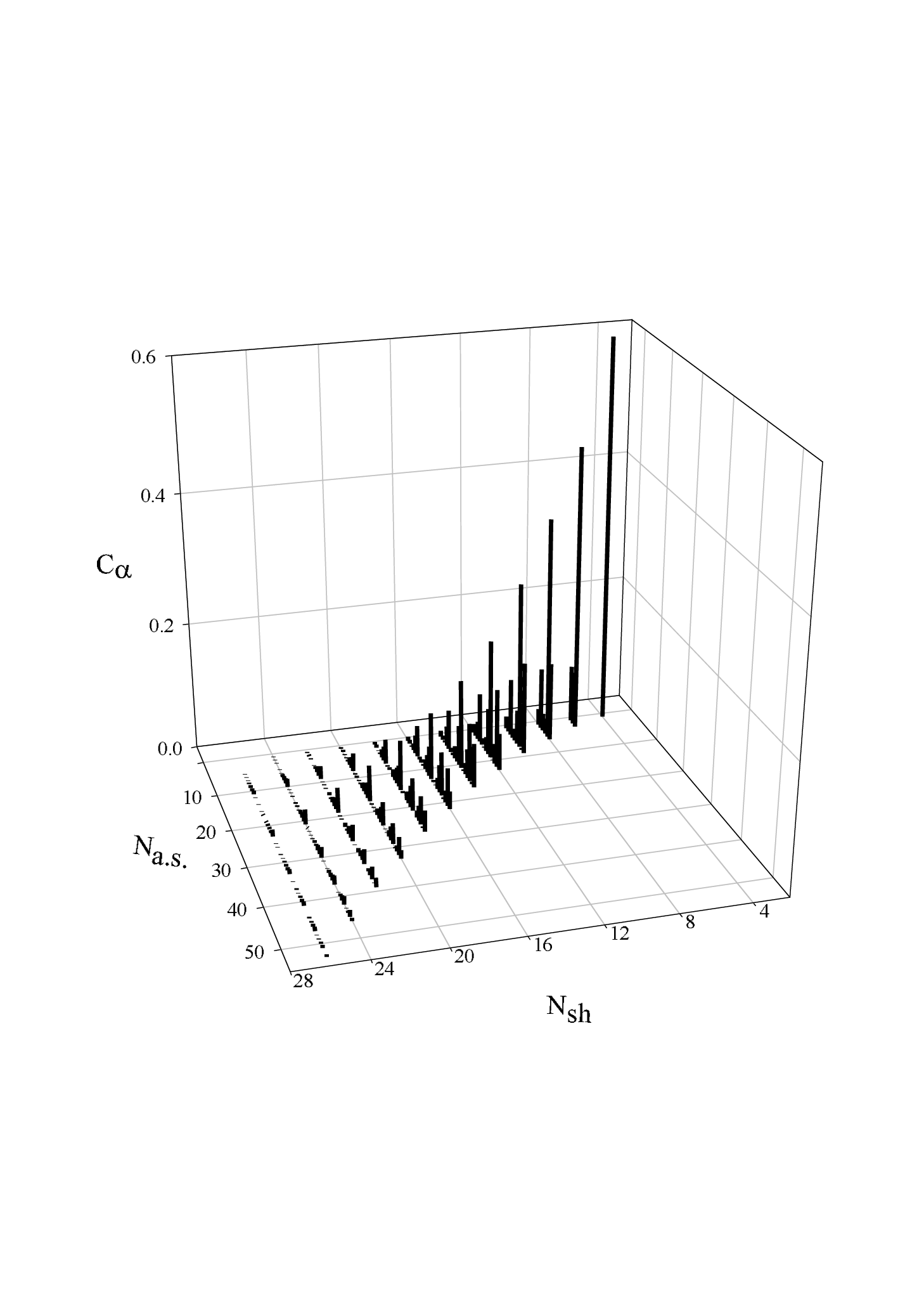}}
\caption{Wave function of the $^{8}$He
ground state in oscillator representation.}
\label{fig:He8_wfun}
\end{figure}

In Fig. \ref{fig:He8_wfun}, we display the wave function of the $^{8}$He
ground state, more exactly, the coefficients $C_{\alpha }$ of expansion over
Pauli-allowed states. Two labels $N_{sh}$ and $N_{a.s}$ are used to classify
Pauli-allowed functions ($\alpha =\left\{ N_{sh},N_{a.s}\right\} $). The
first label $N_{sh}$ numerates oscillator shells and the second one, $%
N_{a.s.}$, numerates Pauli-allowed states of a given oscillator shell. \ The
expansion coefficients $C_{\alpha }$ were determined in the $SU(3)$-basis,
where the indices $\left( \lambda \mu \right) $ are good quantum numbers
after eliminating Pauli-forbidden states. The detailed analysis shows that
the main contribution (around 80\%) to the wave function comes from the
basis states with $\mu =2$, while the basis states with $\mu =0$ give only
9\%. \ Note that the former states in $^{6}$He (see \cite{kn:Vasil96}, \cite
{kn:Vasil97}) were a dominated subspace with the contribution of more than
93\%.

\begin{figure}[tbp]
\centerline{\psfig{figure=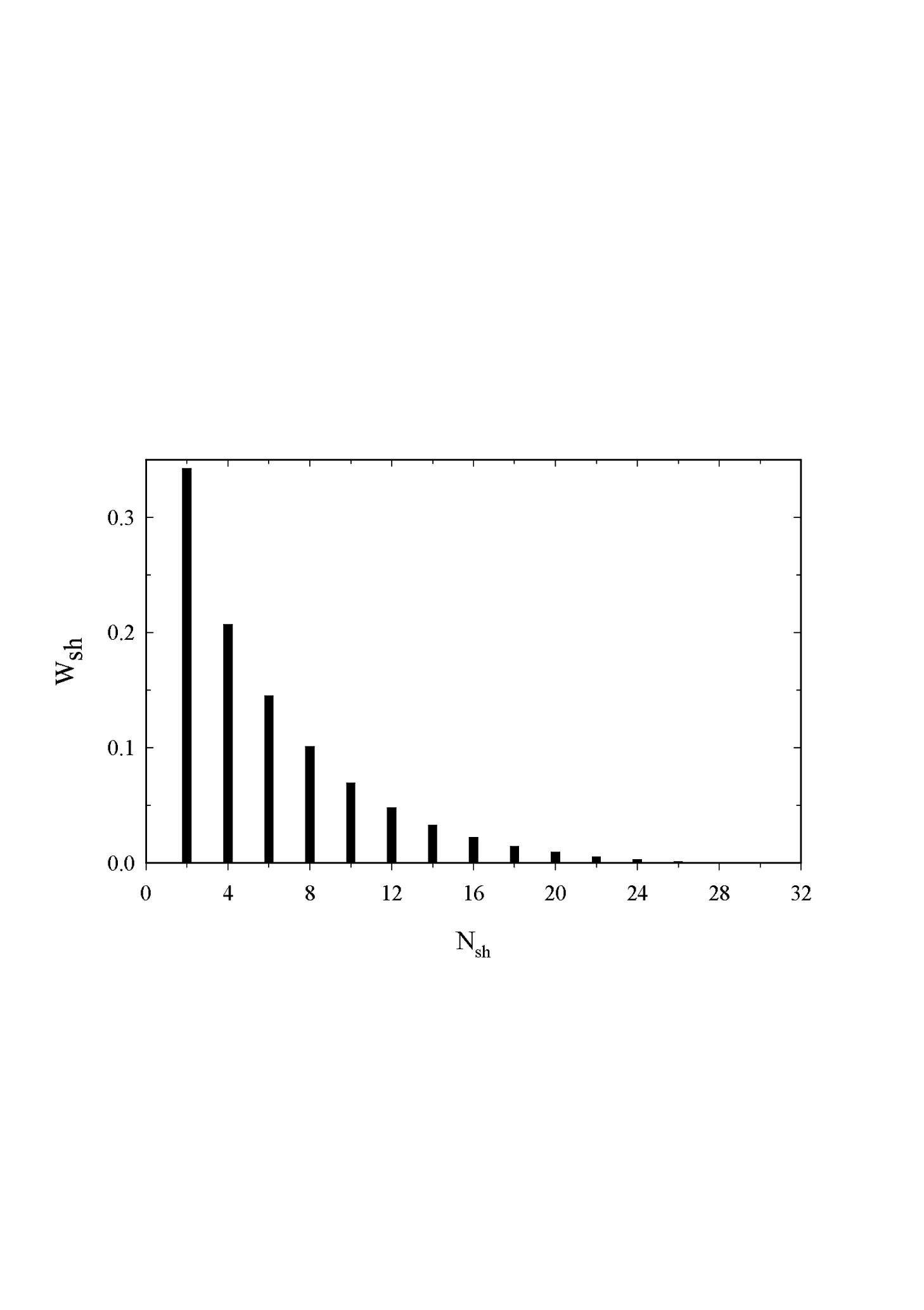}}
\caption{Contribution of different
oscillator shells $N$ to the wave function of the $^{8}$He ground state.}
\label{fig:He8_sh_weights}
\end{figure}

It is seen from Fig.\ref{fig:He8_wfun} and more clearly from Fig. \ref
{fig:He8_sh_weights} (where weights of different oscillator shells are
displayed) that the main contribution comes from\ the lowest oscillator
shells, however the contribution of shells with large $N$ is also
noticeable. It indicates a substantial clusterization of the nuclei, i.e.,
for a large amount of time, valent neutrons move far from the $\alpha $%
-particle, making a neutron halo.

\bigskip

In order to obtain additional information on the role of different subspaces
of the total space of oscillator functions, we impose various restrictions
on the quantum numbers of basis states. First, for the bioscillator basis we
took a subspace \ with the maximal value of partial angular momenta ($%
l=l_{1}=l_{2}$) $l=0$, $l=2$ and $l=4$. It was made for both $Y$- and $T$%
-trees of Jacobi vectors. For the $SU(3)$-basis, we used only $T$-tree and
the restriction was imposed on the maximal value of $\mu =0$, $2$ and $4$.
Results of such calculations are presented in Table \ref{tab:Energy}. One
can see that subspace $l_{1}=l_{2}\leq 2$ for $Y$-tree of the bioscillator
basis is the most optimal part of the total basis, because 54\% of the total
basis (or 219 functions) gives the ground state energy very close to the
''exact'' value.
\bigskip \renewcommand{\baselinestretch}{2.0} 
\begin{table}[t]
\caption{Ground state energy of $^{8}He$ counted from the threshold $%
^{4}He+^{2}n+^{2}n$. }
\label{tab:Energy}
\par
\begin{center}
\begin{tabular}{|c|c|c|c|c|}
\hline
Basis & Jacobi & Subspace & E, MeV & Number \\ 
& tree &  &  & of functions \\ \hline
BO & T & Total & $-2.065$ & $399$ \\ 
&  & $l=0$ & $-1.832$ & $118$ \\ 
&  & $l\leq2$ & $-2.047$ & $219$ \\ 
&  & $l\leq4$ & $-2.064$ & $293$ \\ \hline
BO & Y & Total & $-2.065$ & $399$ \\ 
&  & $l=0$ & $-1.839$ & $118$ \\ 
&  & $l\leq2$ & $-2.065$ & $219$ \\ \hline
SU(3) & T & Total & $-2.065$ & $399$ \\ 
&  & $\mu=0$ & $6.341$ & $91$ \\ 
&  & $\mu\leq2$ & $0.172$ & $196$ \\ 
&  & $\mu\leq4$ & $-1.335$ & $274$ \\ \hline
\end{tabular}
\end{center}
\end{table}

{\bf Effects of the Pauli principle}. To understand the role of the Pauli
principle in a three-cluster system, we investigate the contribution of
different Pauli-allowed states to the wave function of ground state. Each of
Pauli-allowed states, being an eigenfunction of \ the antisymmetrization
operator, can be marked (characterized) by corresponding eigenvalue of this
operator. As we mentioned above, the antisymmetrization operator overlaps
only those functions which obey the relation $N_{1}+N_{2}=N_{1}^{\prime
}+N_{2}^{\prime }$, i.e., basis functions of the same oscillator shell.
Analysis of the eigenfunctions shows, that the diagonalization of the matrix 
$||<n\left| \widehat{A}\right| n^{\prime }>||$ reveals states with definite
eigenvalues of the antisymmetrization operator for a two-cluster subsystem.
This means, in particular, that the Pauli-forbidden state of three-cluster
system is a state when at least one pair of clusters is in Pauli-forbidden
state. For example, for the two-cluster subsystem $\alpha +^{2}n$, the
oscillator functions with the number of oscillator quanta (along the
inter-cluster coordinate) $N=0$ and $N=1$ are Pauli-forbidden states. In the
subsystem $^{2}n+^{2}n$, where symmetry of the subsystem allows only even
functions, we have only one forbidden state with $N=0$. As for Pauli-allowed
states for three-cluster system, they describe the states of the system,
when all pairs of two-cluster subsystems are out of the Pauli-forbidden
region.

\begin{figure}[tbp]
\centerline{\psfig{figure=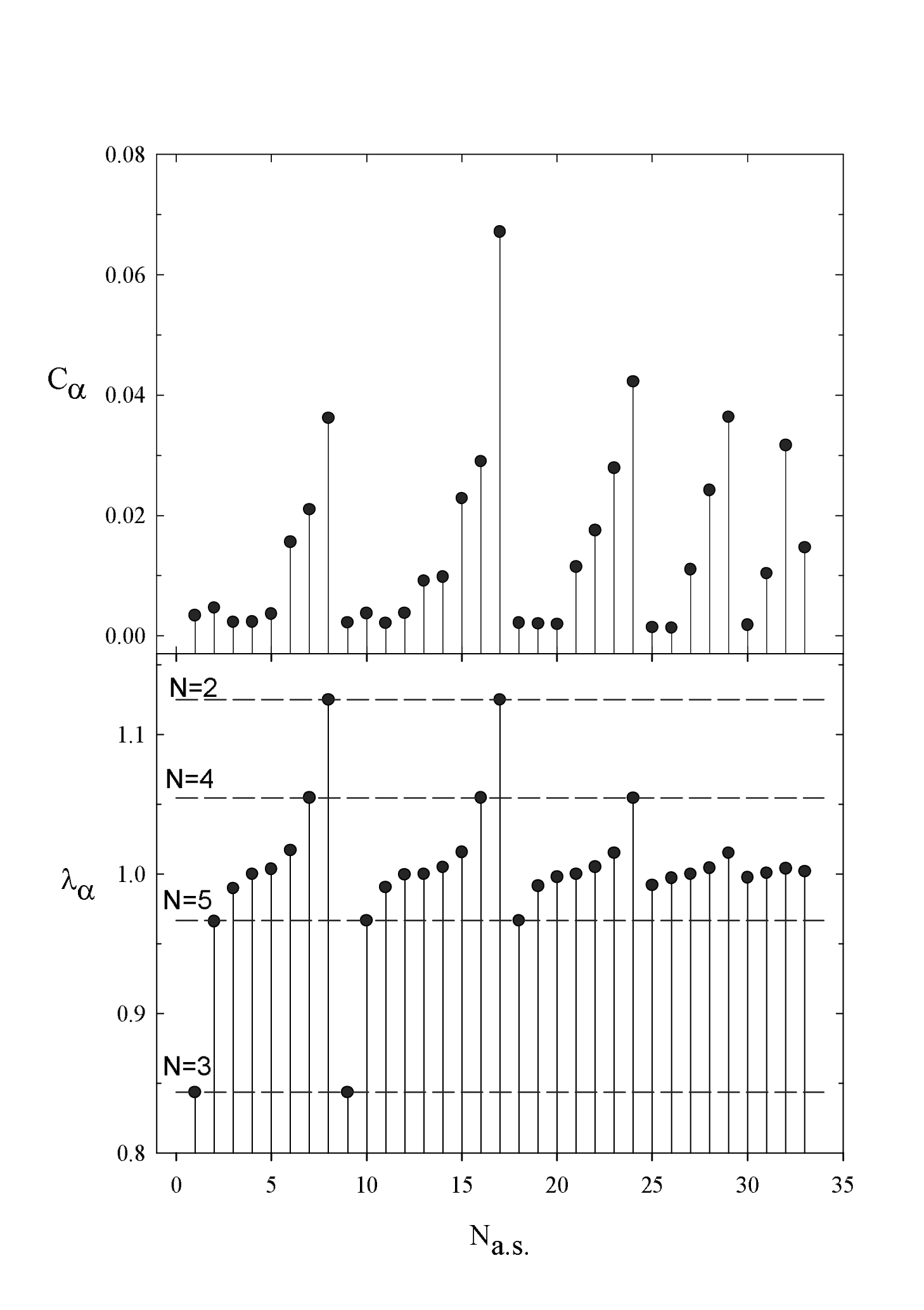}}
\caption{Eigenvalues of the
antisymmetrization operator $\protect\lambda_{N_{a.s.}}$ and expansion
coefficients $C_{\protect\alpha }$ for the oscillator shell $N=20$.}
\label{fig:C_vs_E_n20}
\end{figure}

To prove these statements, we consider eigenvalues of the antisymmetrization
operator (which we denote by $\lambda _{\alpha }$) for Pauli-allowed states
and expansion coefficients $C_{\alpha }$ over these states for oscillator
shell $N_{sh}=20$. These quantities, obtained in the $SU(3)$ basis, are
displayed in Fig. \ref{fig:C_vs_E_n20}. Seven first functions correspond to
the $SU(3)$ irreducible representation $\left( \lambda \mu \right) =\left(
20,0\right) $, next eight functions belong to the $SU(3)$ irreducible
representation $\left( \lambda \mu \right) =\left( 16,2\right) $ and so on.
Last function has the $SU(3)$ symmetry $\left( \lambda \mu \right) =\left(
0,10\right) $. In this Figure by dashed horizontal lines we indicate the
eigenvalues of the antisymmetrization operator for two-cluster subsystem $%
\alpha +^{2}n$. Note that corresponding values for the subsystem $%
^{2}n+^{2}n $ equal 1. One can see indeed, that some eigenvalues of the
operator $\widehat{A}$ for three-cluster system coincide with the
eigenvalues of this operator for the subsystem $\alpha +^{2}n$. Besides, one
notices that such states, corresponding to the even oscillator quanta $N=2$
and $N=4$ in the $\alpha +^{2}n$ subsystem, play dominant role in the ground
state of $^{8}$He.

{\bf RMS\ radii}. In Table \ref{tab:RMSR}, we compare the calculated mass,
neutron and proton root-mean-square (RMS) radii with available experimental
data. The theoretical values of RMS radii are a little larger than
experimental ones which we took from \cite{Alkhazov91} and \cite{kn:Tani88}.
\bigskip \renewcommand{\baselinestretch}{2.0} 
\begin{table}[t]
\caption{Root mean square radii for the ground state of $^{8}He$. }
\label{tab:RMSR}
\par
\begin{center}
\begin{tabular}{|c|c|c|c|}
\hline
RMS, fm & Theory & Experiment \cite{Alkhazov91} & Experiment \cite{kn:Tani88}
\\ \hline
RMSm & $2.73$ & $2.37 \pm 0.18$ & $2.52 \pm 0.03$ \\ 
RMSp & $2.08$ & $1.89 \pm 0.17$ & $2.15 \pm 0.02$ \\ 
RMSn & $2.91$ & $2.50 \pm 0.19$ & $2.64 \pm 0.03$ \\ 
RMSn-RMSp & $0.84$ & $0.61$ & $0.49$ \\ \hline
\end{tabular}
\end{center}
\end{table}
This is perhaps because the calculated binding energy is a little less than
the experimental one. But the present model correctly reproduces the general
picture of $^{8}$He. One sees that radius of neutron matter is larger then
the one of proton matter. The difference of these radii is 0.84 fm. These
results also indicate the existence of a neutron halo in the nucleus.

In Table \ref{tab:RMSR_comp}, we collected the mass, neutron and proton RMS
radii, obtained with different theoretical methods: AV RGM (present
calculations), the Refined Resonating Group Method (RRGM)\cite{hofman97},
and Multi-configuration Shell Model \cite{Navratil+barrett98}.
\begin{table}[]
\caption{Root mean square radii of $^8He$, obtained by different methods. }
\label{tab:RMSR_comp}
\par
\begin{center}
\begin{tabular}{|c|c|c|c|}
\hline
RMS, fm & AV RGM & RRGM \cite{hofman97} & Shell Model \cite
{Navratil+barrett98} \\ \hline
RMSm & $2.73$ & $2.41$ &  \\ 
RMSp & $2.08$ & $1.71$ & $1.684$ \\ 
RMSn & $2.91$ &  &  \\ \hline
\end{tabular}
\end{center}
\end{table}

{\bf Shape of three-cluster system}. Having calculated the coefficients $%
C_{n}$, a wave function in the oscillator representation, we thus obtain the
wave function for relative motion of the three-cluster system: 
\begin{equation}
\Phi ({\bf q}_{1},{\bf q}_{2})=\sum_{n}C_{n}\phi _{n}({\bf q}_{1},{\bf q}%
_{2}).  \label{eq:7}
\end{equation}

By using (\ref{eq:7}), we can evaluate mean distance between clusters. For
this aim, one has to calculate the following quantities: 
\begin{eqnarray*}
Q_{1}^{2} &=&\int d{\bf q}_{1}d{\bf q}_{2}~\Phi ^{\ast }({\bf q}_{1},{\bf q}%
_{2})~{\bf q}_{1}^{2}~\Phi ({\bf q}_{1},{\bf q}_{2}), \\
Q_{2}^{2} &=&\int d{\bf q}_{1}d{\bf q}_{2}~\Phi ^{\ast }({\bf q}_{1},{\bf q}%
_{2})~{\bf q}_{2}^{2}~\Phi ({\bf q}_{1},{\bf q}_{2}),
\end{eqnarray*}
which, within the regard for the normalization of\ the wave function and
definition of Jacobi coordinates, define the sought parameters. For
instance, for $T$-tree, the mean value of $q_{1}$ is connected with the base
of an isosceles triangle and the mean value of $q_{2}$ is connected with its
height.

The mean distance between two dineutrons turns out to be 2.33 fm and the
mean distance between $\alpha $-particle and the center of mass of two
dineutrons is 1.42 fm. Thus, in $^{8}$He, three clusters form an isosceles,
almost rectangular triangle with $\alpha $-particle at the vertex of the
right angle.

\begin{figure}[tbp]
\centerline{\psfig{figure=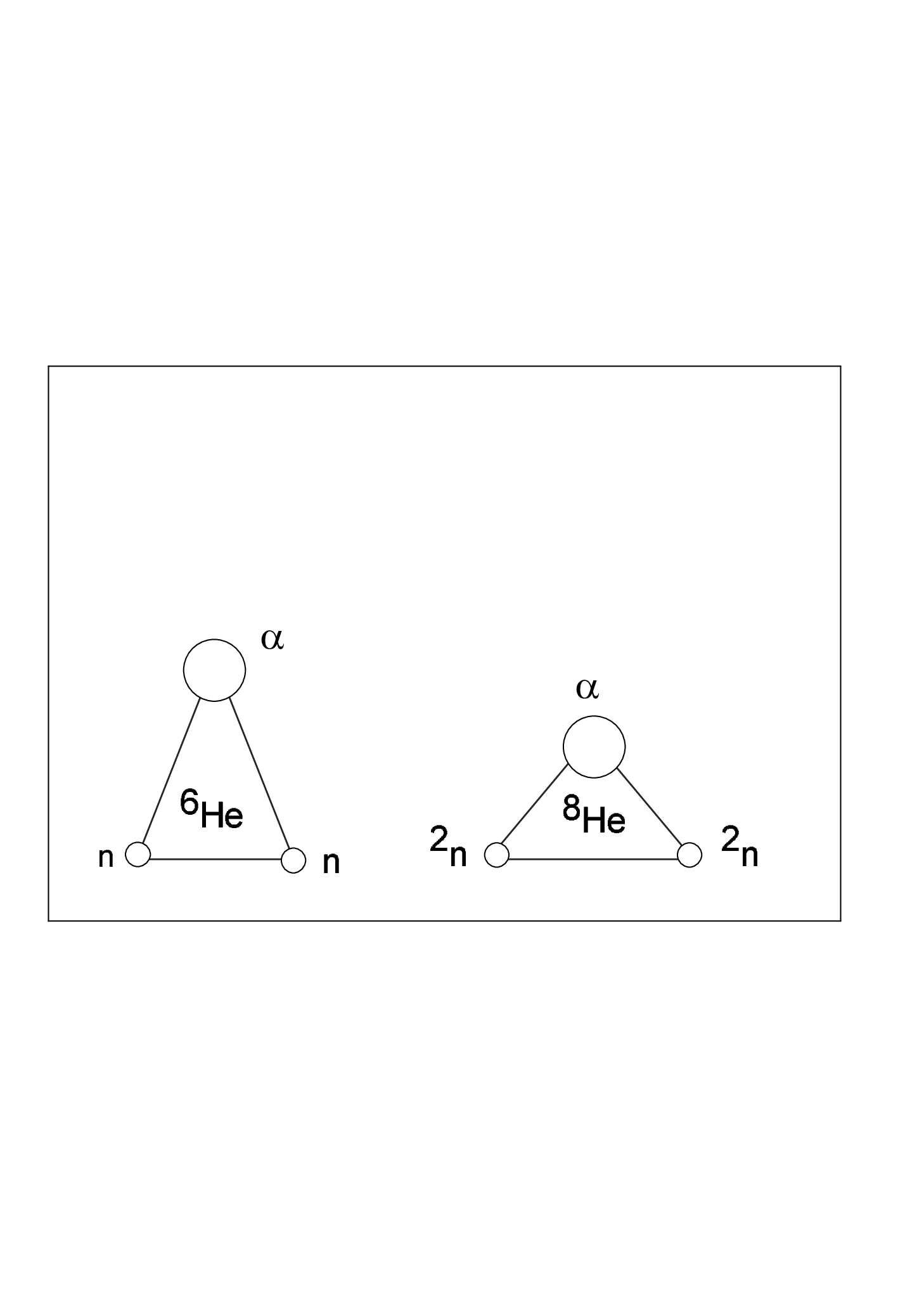}}
\caption{Shape of the triangles which
are composed by three clusters in $^{8}$He and $^{6}$He.}
\label{fig:8He_6He_triangl}
\end{figure}

Note, that the situation is somewhat different for $^{6}$He. Clusters form
an acute-angled triangle. Two valent neutrons in the presence of $\alpha $%
-particle make a subsystem with the RMS radius equal to 2.52 fm which is
less than the RMS radius of a free deuteron (2.69 fm) calculated with the
same potential and the same number of basis functions. These triangles are
displayed in Fig. \ref{fig:8He_6He_triangl}.

The difference in the geometry of cluster's disposition in $^{6}$He and $%
^{8} $He is more likely connected with the Pauli principle. There is an
effective repulsion between two dineutrons, arising from the Pauli
principle, \ which strives to place dineutrons on different sides the $%
\alpha $-particle. Contrary to the case of $^{8}$He, valent neutrons with
opposite orientations of spins may unite in a rather compact subsystem in $%
^{6}$He (due to the presence of $\alpha $-particle).

\begin{figure}[tbp]
\centerline{\psfig{figure=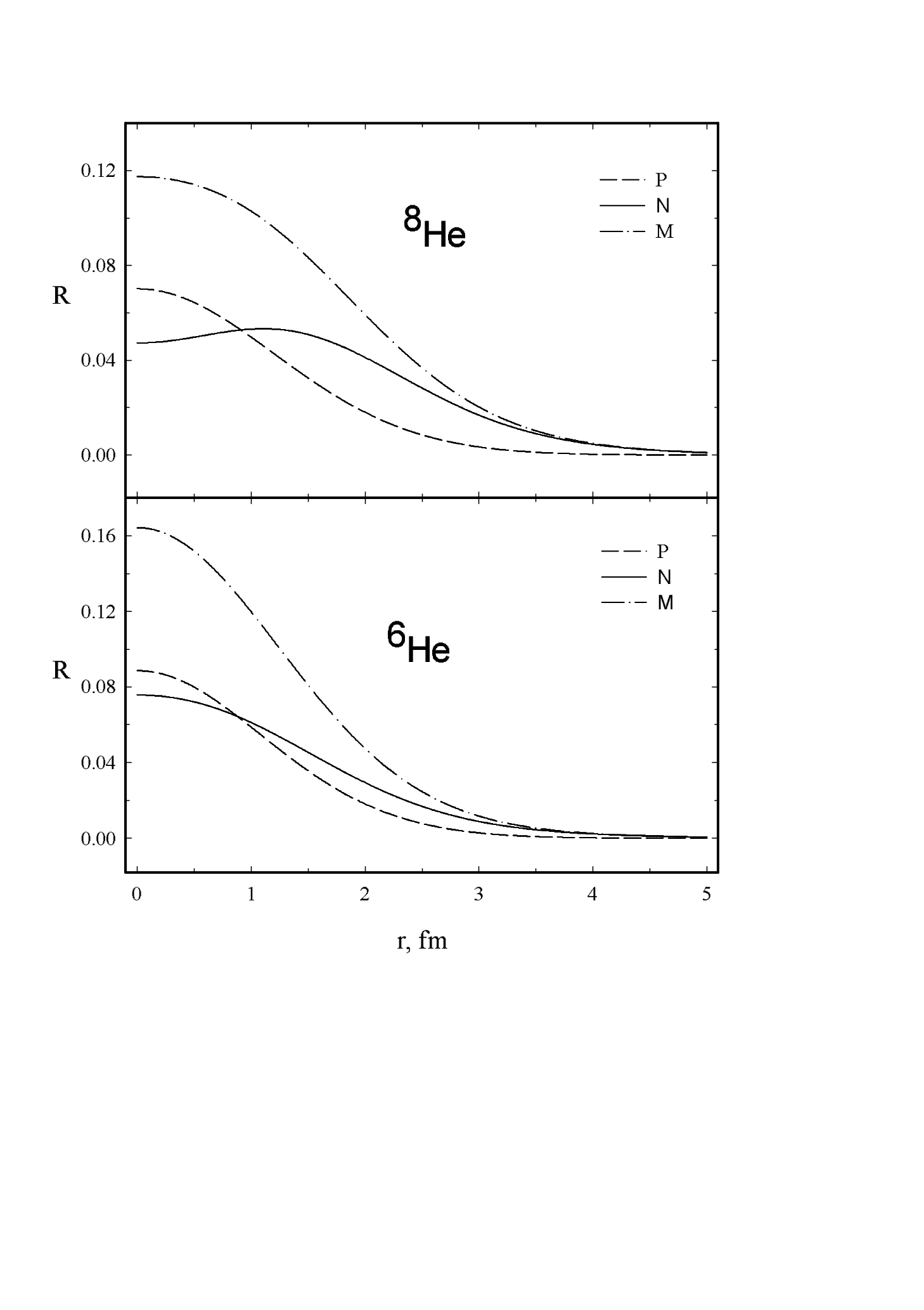}}
\caption{Proton, neutron and mass
density distributions in $^{8}$He and $^{6}$He.}
\label{fig:6He_8He_densities}
\end{figure}

{\bf Density distribution}. Proton, neutron and mass density distributions
also confirm the existence of a neutron halo in $^{8}$He. As seen from Fig. 
\ref{fig:6He_8He_densities}, where we display the proton, neutron and mass

density distributions for both $^{8}$He and $^{6}$He, the size of a neutron
cloud is substantially larger than the size of a proton cloud. Besides, main
part of neutrons in $^{8}$He move on the surface of the nucleus. One sees
the depression in the neutron density distribution at small values of the
coordinate $r$. This is due to the Pauli principle, which makes four
neutrons (united in two dineutrons) move at a relatively large distance from
the $\alpha $-particle.

\section{Conclusion}

In this paper, we have investigated the ground state properties of $^{8}$He
within the three-cluster microscopic model. The three-cluster configuration $%
^{4}$He$+^{2}n+^{2}n$ was used to simulate the dynamics of the eight-nucleon
system. The model suggested describes reasonably well parameters of the
ground state: binding energy, mass, proton and neutron root-mean-square
radii. The analysis of the system shows, that valent neutrons move at a
large distance from $\alpha $-particle, forming a neutron halo in $^{8}$He.

\end{document}